\documentclass[letterpaper]{article} % DO NOT CHANGE THIS
\usepackage{aaai25}  % DO NOT CHANGE THIS
\usepackage{times}  % DO NOT CHANGE THIS
\usepackage{helvet}  % DO NOT CHANGE THIS
\usepackage{courier}  % DO NOT CHANGE THIS
\usepackage[hyphens]{url}  % DO NOT CHANGE THIS
\usepackage{graphicx} % DO NOT CHANGE THIS
\usepackage{natbib}  % DO NOT CHANGE THIS AND DO NOT ADD ANY OPTIONS TO IT
\usepackage{caption} % DO NOT CHANGE THIS AND DO NOT ADD ANY OPTIONS TO IT
\usepackage{algorithm}
\usepackage{algorithmic}
\usepackage{amsmath}
\usepackage{booktabs}
\usepackage{graphicx} 
\usepackage{multirow}
\usepackage{tabularx}
\usepackage{url}
\usepackage{subcaption}
\usepackage{amsmath}
\usepackage{amssymb}

\usepackage{newfloat}
\usepackage{listings}
\DeclareCaptionStyle{ruled}{labelfont=normalfont,labelsep=colon,strut=off} % DO NOT CHANGE THIS
\floatstyle{ruled}
\newfloat{listing}{tb}{lst}{}
\floatname{listing}{Listing}
\title{CodeVision: Detecting LLM-Generated Code Using 2D Token Probability Maps and Vision Models
}
\author {
     Zhenyu Xu, 
    Victor S. Sheng
}
\affiliations {
    Department of Computer Science, Texas Tech University\\
    zhenxu@ttu.edu, victor.sheng@ttu.edu
}

\usepackage{bibentry}
\begin{document}

\maketitle

\begin{abstract}
The rise of large language models (LLMs) like ChatGPT has significantly improved automated code generation, enhancing software development efficiency. However, this introduces challenges in academia, particularly in distinguishing between human-written and LLM-generated code, which complicates issues of academic integrity. Existing detection methods, such as pre-trained models and watermarking, face limitations in adaptability and computational efficiency. In this paper, we propose a novel detection method using 2D token probability maps combined with vision models, preserving spatial code structures such as indentation and brackets. By transforming code into log probability matrices and applying vision models like Vision Transformers (ViT) and ResNet, we capture both content and structure for more accurate detection. Our method shows robustness across multiple programming languages and improves upon traditional detectors, offering a scalable and computationally efficient solution for identifying LLM-generated code.

% multi-language detection/zero-shot, low local computation cost and fast detection.
\end{abstract}

% \begin{IEEEkeywords}
% \end{IEEEkeywords}

\section{Introduction}

Recent advances in LLMs have dramatically improved their capabilities in understanding and generating code. Commercial models like ChatGPT \cite{openai2023chatgpt} and Codex \cite{chen2021evaluating}, along with powerful tools from the open-source community such as CodeLlama \cite{meta2023codellama}, CodeQwen \cite{alibaba2023codeqwen}, StarCoder2 \cite{bigcode2023starcoder}, and DeepSeek \cite{deeplake2023deepseek}, exhibit near-expert levels of programming proficiency. These models not only enhance the efficiency of software development but also democratize programming skills, making it easier for non-experts to engage in technology creation and problem solving. However, these technological advances present unprecedented challenges in computer science education, particularly in terms of academic integrity. As LLMs become increasingly integrated into educational settings, they can inadvertently encourage dependency, where students might rely on these tools via commercial APIs or open-source platforms to effortlessly produce high-quality code for assignments and exams. This raises concerns because the sophisticated nature of LLM outputs, which closely mimic human coding styles, complicates the differentiation between LLM-generated and student-written code \cite{xu2024chatgpt}, rendering traditional plagiarism detection ineffective. This situation underscores the pressing need for reliable and efficient tools to detect LLM-generated code.

Recent research has introduced various methodologies for detecting model-generated code, reflecting significant advancements in this area. A prominent approach involves a pre-trained model detection system \cite{nguyen2023snippet,oedingen2024chatgpt} that utilizes code as input to predict its origin, employing pre-trained feature embedding models alongside supervised learning algorithms . This system transforms code snippets into high-dimensional vector spaces and applies classifiers to ascertain whether the code was human-authored or LLM-generated. Concurrently, watermarking techniques \cite{lee2023watermarking,guan2024codeip} have been developed to embed unique identifiers directly into the outputs of Large Language Models, with some techniques utilizing grammar-guided strategies to embed multi-bit watermarks, thereby enriching the informational content of the generated code. Additionally, perturbation methods \cite{, xu2024detecting,shi2024between, ye2024uncovering} from DetectGPT \cite{mitchell2023detectgpt} strategically insert or replace stylistic tokens, such as code tokens, whitespace, and newline characters, and rewrite segments to highlight the distinct patterns between machine-generated and human-authored code, facilitating accurate predictions regarding the origin of the code .

Current approaches for detecting LLM-generated code face significant limitations. Pre-trained model methods, which rely on input token embeddings, are susceptible to becoming outdated as language models evolve, and struggle with scaling issues, as evidenced by OpenAI's GPT-2 detector study \cite{openai2019gpt2} showing decreased efficacy of smaller detectors against larger models' outputs. Moreover, these methods often require specific fine-tuning for each programming language, limiting their generalizability and increasing the effort needed to maintain effective detection across multiple languages. Wang et al. \cite{wang2023evaluating} demonstrated that these detectors perform poorly when applied to code-related tasks such as code summarization and generation. Watermarking techniques, while promising, may not be optimal for code detection. Suresh et al. \cite{suresh2024watermarking} demonstrated that semantic-preserving modifications can easily circumvent watermark detection for LLM-generated code. Additionally, applying text watermarking techniques \cite{kirchenbauer2023watermark} to code can degrade both code quality and detectability. The DetectGPT perturbation method, although innovative, incurs substantial computational costs and processing time. Bao et al. \cite{bao2023fast} noted that these approaches necessitate approximately 100 model calls or API interactions per detection attempt, rendering them impractical for large-scale or real-time applications. 

To address the aforementioned challenges in existing methods for detecting LLM-generated code, we propose an approach that leverages the unique spatial characteristics of code and utilizes vision models. We use the OpenAI API to calculate log probabilities for each code token, maintaining the original 2D structure of the program. This preserves important spatial features such as indentation, brackets, and line breaks, which are crucial for understanding code structure. We then input this 2D representation of token probabilities into vision models like Vision Transformers \cite{dosovitskiy2020image} and ResNet \cite{he2016deep}. These models are well-suited to capture spatial patterns and long-distance dependencies in code that traditional sequence models might miss. By treating code as a 2D image of probabilities, we enable our models to learn from both the content and the structure of the code simultaneously. This approach is designed to be more robust to changes in language models and coding styles, as it focuses on probability distributions rather than specific tokens. 

Our main contributions are:

\begin{itemize} \item We propose a novel method that converts code into 2D token probability maps and applies vision models, leveraging spatial code features like indentation and brackets for more accurate detection of LLM-generated code. \item Our approach generalizes across multiple programming languages without language-specific fine-tuning, addressing the limitations of previous methods. \item We provide an efficient, real-time detection framework using lightweight vision models, minimizing computational overhead and improving practical applicability in educational and professional settings. \end{itemize}

\section{Related work}
In this section, we review the existing methods for LLM-generated code detection, focusing on approaches such as pre-trained models, watermarking, and perturbation-based techniques.
\subsection{Pre-trained Model}
Nguyen et al. \cite{nguyen2023snippet} introduced GPTSniffer, a tool based on CodeBERT, to identify ChatGPT-generated code. They collected data from GitHub and ChatGPT, extracted features, tokenized and encoded the data using CodeBERT, and trained various classifiers. Oedingen et al. \cite{oedingen2024chatgpt} proposed a similar approach using pre-trained models like CodeBERT. They embedded code snippets using models like TF-IDF and Word2Vec, followed by training classifiers such as SVMs and deep neural networks.

\subsection{Watermark}
SWEET \cite{lee2023watermarking} is a selective watermarking method for code generation. It only watermarks high-entropy tokens, balancing detection effectiveness with code quality. This approach outperformed existing methods in identifying machine-generated code while maintaining code functionality. Signal-based watermarking approaches like FreqMark and signal watermarking \cite{xu2024freqmark, xu2024signal} also demonstrated robust detection capabilities. Beyond Binary Classification \cite{xu2024beyond} introduces a customizable watermarking technique embedding metadata like model origin, generation date, and user identifiers. CodeIP \cite{guan2024codeip} is a grammar-guided multi-bit watermarking technique for code generation by large language models. It embeds watermarks during the code generation process by manipulating the probability distribution of the model's vocabulary. The method incorporates a type predictor to forecast the grammatical type of the next token, ensuring syntactical and semantic correctness of the generated code.

\subsection{Perturbation}
The perturbation method was first proposed and applied by DetectGPT \cite{mitchell2023detectgpt} for machine-generated text detection. Yang et al. \cite{yang2023zeroshot} developed a zero-shot detection method using LLMs to rewrite code and detect machine-generated content by comparing the original and rewritten code's similarity. Xu and Sheng \cite{xu2024detecting} introduced AIGCode Detector, which uses perplexity measures from large language models to detect LLM-generated code, employing targeted mask perturbation and comprehensive scoring. Shi et al. \cite{shi2024between} analyzed differences between machine and human-written code, proposing the DetectCodeGPT method that detects machine-generated code by perturbing stylistic tokens like whitespace. Ye et al. \cite{ye2024uncovering} created a zero-shot synthetic code detector based on the similarity between code and its LLM-rewritten variants, using self-supervised contrastive learning to evaluate the method on synthetic code detection benchmarks.

\section{Approach}

Our approach transforms the task of detecting LLM-generated code into a 2D matrix classification problem by converting code snippets into log probability matrices and using vision models to capture spatial features and make predictions. This method leverages the unique structural characteristics of code and the power of vision models to detect subtle patterns in probability distributions. Key advantages include the preservation of code structure, allowing the model to learn from both the content and layout of the code simultaneously, language agnosticism since we work with log probabilities rather than raw tokens, enabling generalization across programming languages, and robustness to model evolution, as our approach can adapt to advancements in code generation capabilities by using the latest language models to compute log probabilities.

\subsection{Converting Code to Log Probability Matrix}

The first step in our approach is converting code snippets into log probability matrices, preserving the spatial structure of the code, including key features such as indentation, brackets, and line breaks. This process consists of three stages:

\subsubsection{Tokenization} We tokenize the input code snippet using the \texttt{cl100k\_base} encoding, which is optimized for code-like content. This ensures that code-specific tokens, such as symbols and keywords, are accurately recognized.

\subsubsection{Log Probability Calculation} For each token in the sequence, we compute its log probability using a large language model via the OpenAI API. The process involves iterating through the token sequence, sending each token along with the preceding ones as a prompt to the LLM, and retrieving the log probabilities for the top 10 most likely next tokens. We then extract the log probability corresponding to the actual next token in the sequence.

\subsubsection{2D Matrix Construction} The computed log probabilities are arranged into a 2D matrix that mirrors the structure of the original code. We initialize empty matrices for both log probabilities and tokens. Each token and its corresponding log probability are processed sequentially. When a newline character is encountered, it signifies the end of a line, and the current line's log probabilities and tokens are appended to their respective matrices. This continues until all tokens are processed, ensuring that the final matrix accurately reflects the structure of the code, including indentation and line breaks. Empty spaces are filled with a placeholder value (e.g., -100) to maintain the spatial layout. The resulting log probability matrix $M$ has dimensions $n \times m$, where $n$ is the number of lines in the code and $m$ is the maximum number of tokens in any line. Each element $M_{ij}$ represents the log probability of the $j$-th token in the $i$-th line of code. 

% \begin{figure}[h]
%     \centering
%     \includegraphics[width=\linewidth]{prob2.png}
%     \caption{logp.}
%     \label{fig:logp}
% \end{figure}

\subsection{Vision Models for Classification}
\begin{figure}[h]
    \centering
    \begin{subfigure}[b]{0.45\linewidth}
        \centering
        \includegraphics[width=\linewidth]{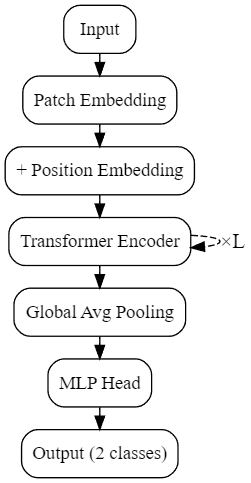}
        \caption{ViT}
        \label{fig:vit}
    \end{subfigure}
    \hfill
    \begin{subfigure}[b]{0.45\linewidth}
        \centering
        \includegraphics[width=\linewidth]{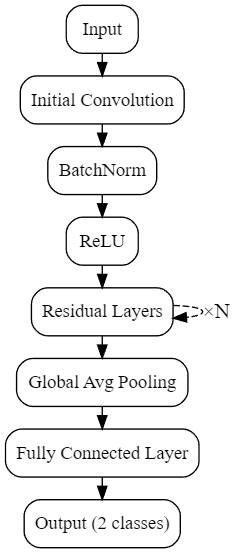}
        \caption{ResNet}
        \label{fig:resnet}
    \end{subfigure}
    \caption{Model Architectures.}
    \label{fig:model}
\end{figure}

Once we have the log probability matrix, we treat it as an image and input it into a vision model for classification. We primarily used a ViT and a modified ResNet for LLM-generated code detection. Figure \ref{fig:model} provides a visual representation of vision model architectures. The ViT processes input patches with a projection layer, uses positional encoding for spatial information, and applies transformer encoder layers with self-attention for feature extraction, followed by a classification head. The ResNet model, adapted for single-channel input, includes convolutional layers, batch normalization, residual blocks, and a fully connected layer for classification. Both models end with global average pooling and are designed to process 2D code representations using either self-attention (ViT) or convolution (ResNet) to capture relevant features.

% digraph ViT {
%     rankdir=TB;
%     node [shape=box, style="rounded,filled", fillcolor=white, fontname="Times New Roman", fontsize=14];
%     edge [fontname="Times New Roman", fontsize=16];
%     graph [nodesep=0.3, ranksep=0.1];

%     input [label="Input"];
%     patch_embed [label="Patch Embedding"];
%     pos_embed [label="+ Position Embedding"];
%     transformer [label="Transformer Encoder", shape=rectangle];
%     pool [label="Global Avg Pooling"];
%     mlp [label="MLP Head"];
%     output [label="Output (2 classes)"];

%     input -> patch_embed -> pos_embed -> transformer;
%     transformer -> pool -> mlp -> output;

%     transformer -> transformer [label="×L", style=dashed, fontname="Times New Roman", constraint=false];
% }
% digraph ResNet {
%     rankdir=TB;
%     node [shape=box, style="rounded,filled", fillcolor=white, fontname="Times New Roman", fontsize=14];
%     edge [fontname="Times New Roman", fontsize=16];
%     graph [nodesep=0.3, ranksep=0.1];

%     input [label="Input"];
%     conv1 [label="Initial Convolution"];
%     bn1 [label="BatchNorm"];
%     relu1 [label="ReLU"];
%     res_layers [label="Residual Layers", shape=rectangle];
%     gap [label="Global Avg Pooling"];
%     fc [label="Fully Connected Layer"];
%     output [label="Output (2 classes)"];

%     input -> conv1 -> bn1 -> relu1 -> res_layers -> gap -> fc -> output;
%     res_layers -> res_layers [label="×N", style=dashed, fontname="Times New Roman", constraint=false];
% }

\subsubsection{Vision Transformer}

The ViT model processes input code as a sequence of patches. The input $\mathbf{X} \in \mathbb{R}^{H \times W \times C}$ is divided into $N$ fixed-size patches and linearly embedded into a dimension $D$:

\[
\mathbf{z}_0 = [\mathbf{x}_p^1 \mathbf{E}; \mathbf{x}_p^2 \mathbf{E}; \dots; \mathbf{x}_p^N \mathbf{E}] + \mathbf{E}_{pos}
\]
where $\mathbf{E} \in \mathbb{R}^{(P^2 \cdot C) \times D}$ is the patch embedding projection and $\mathbf{E}_{pos} \in \mathbb{R}^{N \times D}$ is the position embedding.

The Transformer Encoder consists of $L$ identical layers. Each layer processes the input as follows:
\[
\mathbf{z}'_l = \text{MSA}(\text{LN}(\mathbf{z}_{l-1})) + \mathbf{z}_{l-1}
\]
\[
\mathbf{z}_l = \text{MLP}(\text{LN}(\mathbf{z}'_l)) + \mathbf{z}'_l
\]
where MSA is Multi-Head Self-Attention, LN is Layer Normalization, and MLP is a Multi-Layer Perceptron.

For classification, we apply Global Average Pooling (GAP) and a Multi-Layer Perceptron (MLP):
\[
\mathbf{y} = \text{MLP}(\text{GAP}(\mathbf{z}_L))
\]
where $\mathbf{y} \in \mathbb{R}^2$ represents the output logits for binary classification.

\subsubsection{Modified ResNet}

The ResNet model is adapted for single-channel input $\mathbf{X} \in \mathbb{R}^{H \times W \times 1}$. It starts with an initial convolution layer:
\[
\mathbf{x}_1 = \text{ReLU}(\text{BN}(\text{Conv}_{3\times3}(\mathbf{X})))
\]
where Conv$_{3\times3}$ is a 3×3 convolution, BN is Batch Normalization, and ReLU is the activation function.

Each residual block follows:
\[
\mathbf{x}_{l+1} = \text{ReLU}(\mathbf{x}_l + \mathcal{F}(\mathbf{x}_l, \mathbf{W}_l))
\]
where the residual function $\mathcal{F}$ is defined as:
\[
\mathcal{F}(\mathbf{x}_l, \mathbf{W}_l) = \text{BN}(\text{Conv}_{3\times3}(\text{ReLU}(\text{BN}(\text{Conv}_{3\times3}(\mathbf{x}_l)))))
\]

Global Average Pooling is applied after the final residual block, followed by a Fully Connected (FC) layer for classification:
\[
\mathbf{y} = \text{FC}(\text{GAP}(\mathbf{x}_L))
\]
where $\mathbf{y} \in \mathbb{R}^2$ represents the output logits.

\subsubsection{Loss Function}

Both models are trained using cross-entropy loss:
\[
\mathcal{L} = -\frac{1}{N}\sum_{i=1}^N \left[ y_i \log(\hat{y}_i) + (1-y_i) \log(1-\hat{y}_i) \right]
\]
where $y_i$ is the true label and $\hat{y}_i$ is the predicted probability for the $i$-th sample.

\section{Experimental Setting} In this section, we describe the datasets, baselines, evaluation metrics, research questions, and implementation details for our experiments.

\subsection{Datasets}
We utilized a dataset derived from Project CodeNet \cite{puri2021codenet}, consisting of programming problems in six languages: C, C++, Go, Java, Python, and Ruby. The dataset includes 28,342 code samples, with 15,633 generated by GPT-3.5 and GPT-4, and 12,709 written by humans. Each sample is annotated with a 2D probability matrix, a 1D log probability vector, and a binary label (1 for LLM-generated, 0 for human-written). The data is split into 80\% for training, 10\% for validation, and 10\% for testing, ensuring consistent ratios of LLM-generated to human-written code across all sets.

\subsection{Baselines}
Our evaluation is benchmarked against a diverse range of machine-generated text detection techniques, as well as supervised baselines:

\begin{itemize}
\item \textbf{Logp(x)} \cite{gehrmann2019gltr}: Computes the mean log probability of tokens to assess code authenticity. LLM-generated code typically yields higher values.
\item \textbf{Entropy} \cite{xu2020detecting}: Analyzes the average uncertainty in the model's token predictions. Lower entropy often signifies machine-generated content.
\item \textbf{(Log-) Rank} \cite{ippolito2020automatic}: Calculates the mean (log) position of each token in the model's probability distribution. LLM-generated text generally results in lower average ranks.
\item \textbf{DetectGPT} \cite{mitchell2023detectgpt}: Evaluates the difference in log probabilities between the original code and its perturbed versions to identify machine-generated content.
\item \textbf{DetectLLM} \cite{zhao2023detectllm}: Proposes two novel approaches: LRR (Log-Likelihood Log-Rank Ratio), combining log likelihood and rank, and NPR (Normalized Perturbed log-Rank), an enhancement of DetectGPT's methodology.
\item \textbf{GPT-2 Detector} \cite{solaiman2019release}: Employs a supervised learning approach with a RoBERTa-based model to distinguish between human and AI-authored text.
\item \textbf{GPTZero} \cite{tian2023gptzero}: Utilizes a commercial API trained on output from various large language models to detect LLM-generated content.
\end{itemize}

\subsection{Evaluation Metrics}
We employ three key metrics to evaluate detector performance: \textbf{AUC}, \textbf{FPR}, and \textbf{FNR}. The Area Under the Receiver Operating Characteristic Curve (AUC) measures overall discrimination ability, with scores closer to 1.0 indicating better performance. False Positive Rate (FPR) represents the proportion of human-written code mistakenly flagged as LLM-generated, while False Negative Rate (FNR) shows the fraction of LLM-generated code incorrectly classified as human-written. Together, these metrics provide a comprehensive assessment of detection accuracy and error rates. Additionally, we use \textbf{FLOPs} (Floating Point Operations per Second) to quantify the computational resources required, allowing us to compare the efficiency of different models and methods.

\subsection{Research Questions}
Our experiment addresses three primary research questions:
\subsubsection{RQ 1: Performance Evaluation}
We assess the effectiveness of our model in detecting LLM-generated code by comparing it to established baselines. This includes analyzing performance across different code generators.

\subsubsection{RQ 2: Robustness Against Attacks}
We evaluate the model’s resilience to common evasion techniques, focusing on three types of attacks: code mixing, code translation, and insertion of redundant code snippets.

\subsubsection{RQ 3: Impact of Model Scaling}
We examine how variations in model architecture and input characteristics affect detection performance. This includes analyzing the impact of different parameter scales in ResNet and ViT, investigating the influence of varying code lengths on detection accuracy, and evaluating detection speed for practical applicability.

\subsection{Implementation Details}

Log probabilities for each token were calculated using GPT-3.5-turbo-instruct with a temperature of 0 and top-p set to 1 for deterministic outputs. The dataset was generated in July 2024 via OpenAI's API, using both GPT-3.5-turbo-instruct and GPT-4-turbo. Unless noted otherwise, default parameters (temperature 0.9, top-p 0.95) were applied, similar to the ChatGPT web interface.

We implemented three base models: ResNet and ViT for processing 2D log probability matrices, and Transformer-XL for sequence log probabilities. Transformer-XL was chosen for its ability to handle longer sequences. The ViT model utilizes a 128-dimensional embedding, 3 Transformer encoder layers with 4 attention heads each, and a 256-dimensional MLP. ResNet was modified to a 56-layer convolutional network with 9 basic blocks per stage, across three stages. Transformer-XL has 6 layers, 8 attention heads, a 512-dimensional hidden layer, and a memory length of 50. Binary classification layers were added to all models. Training was conducted with the Adam optimizer (learning rate 1e-4), with early stopping based on validation performance.

Baselines include DetectGPT and NPR (DetectLLM), which use a perturbation-based method masking 15\% of code tokens with a span length of 2. CodeT5p-6B \cite{wang2023codet5plus} was used for mask filling, with temperature set to 1, repeated 100 times. The GPT-2 Detector employs a RoBERTa-base model \cite{liu2019roberta} with 125M parameters.

\section{Main Results} In this section, we present the key findings from our experiments, evaluating the performance of our models across multiple programming languages, assessing their robustness against various attacks, and analyzing time consumption.

\subsection{Performance Evaluation}

We evaluate the performance of our proposed models, ResNet, ViT and TransformerXL, against baselines across six programming languages, as shown in Table \ref{table:p}. Our models demonstrate strong effectiveness in detecting LLM-generated code, along with balanced FPR and FNR values across all languages. ResNet and ViT consistently outperform TransformerXL, underscoring the advantages of using 2D matrix representations for capturing code structures compared to sequence-based models. The highest performance is observed on Python, likely due to more extensive training data, while Ruby shows slightly lower results.

In Table \ref{table:clm}, we compare the performance of various detection methods on Python code generated by different Code LLMs. The evaluated methods include Logp(x), Entropy, Rank, Log Rank, DetectGPT, LRR, NPR, TransformerXL, ResNet, and ViT. Among these, ResNet and ViT achieve the highest performance, followed by DetectGPT and NPR. In contrast, methods like Logp(x) and LRR show weaker performance, particularly on smaller models such as SantaCoder and CodeGeex2. The lower performance on small code LLMs is likely due to the lower quality of code they generate, which may contain errors. These discrepancies cause the calculated log probabilities to diverge from what LLMs typically expect, leading to a decline in AUC scores.

\begin{table*}[!ht]
    \centering
    \footnotesize
    \setlength{\tabcolsep}{6pt} 
    \begin{tabular}{lp{0.35cm}p{0.35cm}p{0.35cm} p{0.35cm}p{0.35cm}p{0.35cm} p{0.35cm}p{0.35cm}p{0.35cm} p{0.35cm}p{0.35cm}p{0.35cm} p{0.35cm}p{0.35cm}p{0.35cm} p{0.35cm}p{0.35cm}p{0.35cm}}
    \toprule
    & \multicolumn{3}{c}{\textbf{C}} & \multicolumn{3}{c}{\textbf{C++}} & \multicolumn{3}{c}{\textbf{GO}} & \multicolumn{3}{c}{\textbf{Java}} & \multicolumn{3}{c}{\textbf{Python}} & \multicolumn{3}{c}{\textbf{Ruby}} \\
    \cmidrule(lr){2-4} \cmidrule(lr){5-7} \cmidrule(lr){8-10} \cmidrule(lr){11-13} \cmidrule(lr){14-16} \cmidrule(lr){17-19}
    \textbf{Detectors} & AUC & FPR & FNR & AUC & FPR & FNR & AUC & FPR & FNR & AUC & FPR & FNR & AUC & FPR & FNR & AUC & FPR & FNR \\
        \midrule
        Logp(x) & 0.57 & 0.86 & 0.09 & 0.69 & 0.83 & 0.00 & 0.38 & 0.96 & 0.03 & 0.56 & 0.89 & 0.08 & 0.44 & 0.91 & 0.38 & 0.34 & 0.97 & 0.25 \\
        Entropy & 0.61 & 0.00 & 0.95 & 0.42 & 0.00 & 0.88 & 0.56 & 0.00 & 1.00 & 0.45 & 0.03 & 0.87 & 0.59 & 0.00 & 1.00 & 0.48 & 0.00 & 0.97 \\
        Rank & 0.70 & 0.08 & 0.87 & 0.51 & 0.14 & 0.88 & 0.45 & 0.00 & 0.99 & 0.57 & 0.00 & 1.00 & 0.48 & 0.15 & 0.92 & 0.52 & 0.00 & 1.00 \\
        Log Rank & 0.63 & 0.00 & 1.00 & 0.62 & 0.13 & 0.92 & 0.37 & 0.00 & 1.00 & 0.58 & 0.07 & 0.88 & 0.50 & 0.10 & 1.00 & 0.64 & 0.10 & 0.86 \\
        
        GPTZero & 0.80 & 0.10 & 0.20 & 0.72 & 0.28 & 0.64 & 0.23 & 0.18 & 1.00 & 0.24 & 0.26 & 0.95 & 0.39 & 0.17 & 1.00 & 0.55 & 0.14 & 1.00 \\
        GPT2 Detector & 0.66 & 0.21 & 0.96 & 0.52 & 0.17 & 1.00 & 0.50 & 0.23 & 0.97 & 0.50 & 0.04 & 1.00 & 0.44 & 0.14 & 0.89 & 0.46 & 0.12 & 0.96 \\

        DetectGPT& 0.90 & 0.11 & 0.11 & 0.86 & 0.14 & 0.00 & 0.66 & 0.14 & 0.04 & 0.92 & 0.15 & 0.06 & 0.97 & 0.24 & 0.11 & 0.66 & 0.17 & 0.05  \\
        
        LRR & 0.53 & 0.89 & 0.12 & 0.64 & 0.97 & 0.12 & 0.26 & 1.00 & 0.12 & 0.44 & 0.92 & 0.14 & 0.37 & 1.00 & 0.50 & 0.32 & 1.00 & 0.25 \\
        
        NPR  & 0.78 & 0.16 & 0.18 & 0.82 & 0.15 & 0.11 & 0.59 & 0.20 & 0.13 &0.94 & 0.18 & 0.09 & 0.91 & 0.29 & 0.17 & 0.51 & 0.27 & 0.10  \\
        
        \midrule
        TransformerXL & 0.89 & 0.18 & 0.15 & 0.87 & 0.21 & 0.14 & 0.93 & 0.17 & 0.19 & 0.90 & 0.19 & 0.12 & 0.92 & 0.15 & 0.16 & 0.85 & 0.23 & 0.17 \\
        ResNet & 0.99 & 0.10 & 0.10 & 0.97 & 0.13 & 0.09 & 0.98 & 0.11 & 0.13 & 0.99 & 0.08 & 0.11 & 0.98 & 0.09 & 0.07 & 0.96 & 0.06 & 0.14 \\
        ViT & 0.98 & 0.14 & 0.13 & 0.98 & 0.12 & 0.15 & 0.95 & 0.17 & 0.13 & 0.96 & 0.15 & 0.10 & 0.99 & 0.11 & 0.12 & 0.94 & 0.18 & 0.11 \\
        \bottomrule
    \end{tabular}
    \caption{Performance of Different Detectors Across Six Programming Languages.}
    \label{table:p}
\end{table*}

\begin{table*}[!ht]
    \centering
    \footnotesize
    \setlength{\tabcolsep}{4pt} 
    \begin{tabular}{lcccccccccc}
    \toprule
    \textbf{Code LLMs} & \textbf{Logp(x)} & \textbf{Entropy} & \textbf{Rank} & \textbf{Log Rank} & \textbf{DetectGPT} & \textbf{LRR} & \textbf{NPR} & \textbf{TransformerXL}& \textbf{ResNet} & \textbf{ViT}\\
    \midrule
    SantaCoder-1.1B   & 0.53 & 0.62 & 0.55 & 0.57 & 0.72 & 0.45 & 0.62 & 0.67 & 0.76 & 0.75 \\
    CodeGeex2-6B      & 0.50 & 0.60 & 0.52 & 0.54 & 0.76 & 0.42 & 0.71 & 0.70 & 0.79 & 0.78 \\
    CodeLlama-7b      & 0.48 & 0.58 & 0.67 & 0.52 & 0.79 & 0.40 & 0.72 & 0.73 & 0.82 & 0.81 \\
    CodeQwen1.5-7B    & 0.47 & 0.71 & 0.49 & 0.51 & 0.81 & 0.39 & 0.77 & 0.75 & 0.84 & 0.83 \\
    StarCoder2-15B    & 0.46 & 0.56 & 0.48 & 0.65 & 0.84 & 0.38 & 0.80 & 0.78 & 0.87 & 0.86 \\
    GPT-3.5-turbo     & 0.44 & 0.59 & 0.48 & 0.50 & 0.89 & 0.54 & 0.89 & 0.85 & 0.98 & 0.96 \\
    GPT-4-turbo       & 0.38 & 0.57 & 0.46 & 0.48 & 0.93 & 0.35 & 0.91 & 0.92 & 0.97 & 0.96 \\
    \bottomrule
    \end{tabular}
    \caption{Comparison of Detection Methods in AUC for Python Codes Across Various Code LLMs.}
    \label{table:clm}
\end{table*}

\subsection{Robustness Against Attacks}

We assess the model's robustness against three attack types: Code Mixing, Code Translation, and Insertion of Redundant Code. In Code Mixing (10\%, 30\%, 50\%), human-written code is inserted into LLM-generated code at various proportions, degrading detection performance as the mix ratio increases. Code Translation, using tools like J2Py, significantly alters the code structure, affecting pattern recognition. Insertion of Redundant Code involves adding non-functional code, renaming identifiers, inserting print statements, or wrapping code in try-catch blocks, all of which introduce structural changes to test the model's resilience. 

Table \ref{tab:attack_results} shows the deviation from baseline performance based on our attack experiments using the ViT model for detecting LLM-generated Python code. Code Translation results in the largest drop in performance, with significant increases in both FPR and FNR. Code Mixing also impacts detection, with greater effects as the mix ratio increases. InsertDeadCode and WrapTryCatch cause moderate degradation in performance, while Rename and InsertPrint have minimal influence on the detection model's accuracy.
\begin{table}[ht]
    \centering
    \footnotesize
    \setlength{\tabcolsep}{8pt} % Adjust column spacing
    \renewcommand{\arraystretch}{1.2} % Adjust row spacing
    \begin{tabular}{lccc}
        \toprule
        \textbf{Attack Type} & \textbf{AUC} & \textbf{FPR} & \textbf{FNR} \\
        \midrule
        Code Mixing (10\%) & 0.02 $\downarrow$ & 0.02 $\uparrow$ & 0.02 $\uparrow$ \\
        Code Mixing (30\%) & 0.06 $\downarrow$ & 0.05 $\uparrow$ & 0.05 $\uparrow$ \\
        Code Mixing (50\%) & 0.11 $\downarrow$ & 0.10 $\uparrow$ & 0.09 $\uparrow$ \\
        \midrule
        Code Translation & 0.25 $\downarrow$ & 0.20 $\uparrow$ & 0.18 $\uparrow$ \\
        \midrule
        InsertDeadCode & 0.09 $\downarrow$ & 0.07 $\uparrow$ & 0.08 $\uparrow$ \\
        Rename & 0.03 $\downarrow$ & 0.01 $\uparrow$ & 0.02 $\uparrow$ \\
        InsertPrint & 0.05 $\downarrow$ & 0.03 $\uparrow$ & 0.04 $\uparrow$ \\
        WrapTryCatch & 0.10 $\downarrow$ & 0.09 $\uparrow$ & 0.09 $\uparrow$ \\
        \midrule
    \end{tabular}
    \caption{Impact of Different Attacks on Detection Performance.}
    \label{tab:attack_results}
\end{table}

\subsection{Impact of Model Scaling}

\begin{figure}[!ht]
    \centering
    \begin{subfigure}[b]{\linewidth}
        \centering
        \includegraphics[width=\linewidth]{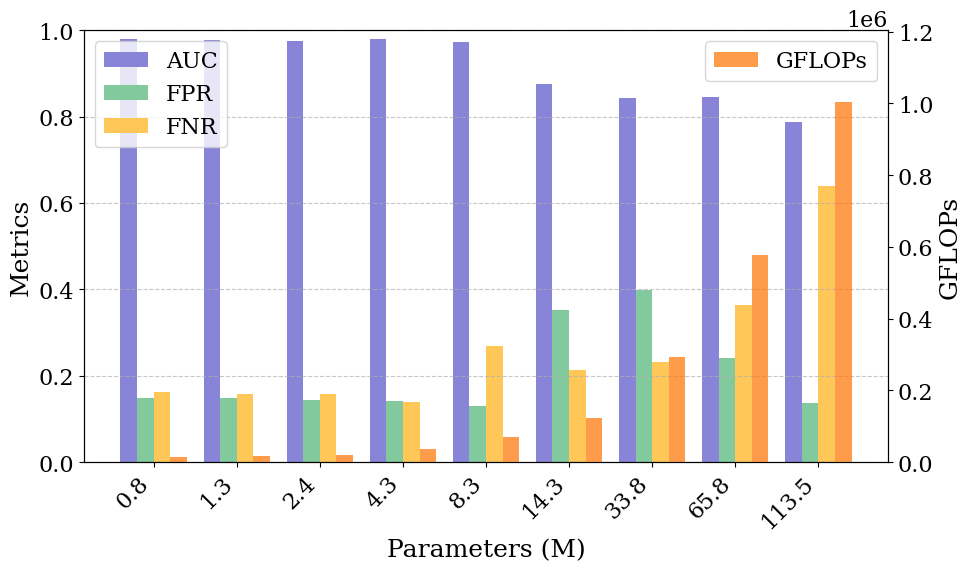}
        \caption{ViT Performance Comparison Across Model Sizes}
        \label{fig:vitp}
    \end{subfigure}
    \hfill
    \begin{subfigure}[b]{\linewidth}
        \centering
        \includegraphics[width=\linewidth]{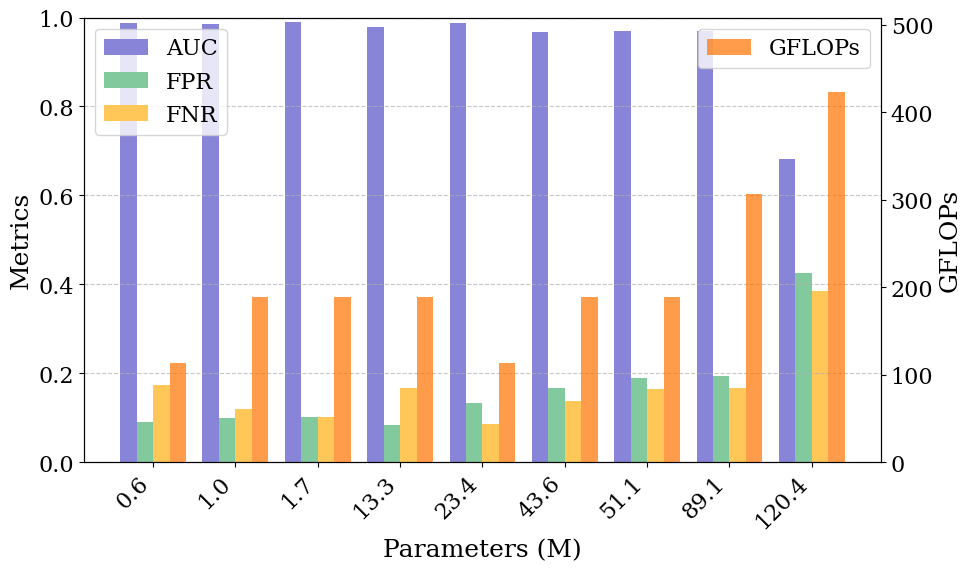}
        \caption{ResNet Performance Comparison Across Model Sizes}
        \label{fig:resnetp}
    \end{subfigure}
    \caption{Performance and FLOPs of ViT and ResNet Models of Scaling Sizes.}
    \label{fig:modelp}
\end{figure}

Figure \ref{fig:modelp} illustrates the performance of ViT and ResNet architectures across various model sizes for LLM-generated code detection. Both ViT and ResNet architectures demonstrate that smaller models (0.5M to 20M parameters) achieve remarkably high AUC scores ($>$0.97). This suggests that the task of detecting LLM-generated code does not necessarily require large model capacities. As model size increases beyond these optimal points, we observe a performance plateau or even decline, despite substantial increases in parameter count and computational requirements (GFLOPs). These findings suggest that LLM-generated code detection may rely more on the identification of specific patterns or features that can be effectively captured by smaller models. The performance decline in larger models could be attributed to overfitting or the learning of redundant features that do not contribute to improved detection accuracy.

Smaller models perform well because they are able to capture the key patterns and structural features necessary for detecting LLM-generated code without overfitting to irrelevant details. Their lower complexity reduces the risk of learning redundant or noise-related features, making them more efficient for this task while maintaining high accuracy. Additionally, smaller models require fewer computational resources, making them ideal for practical deployment in environments like classrooms.

\begin{figure}[!ht]
    \centering
    \includegraphics[width=0.9\linewidth]{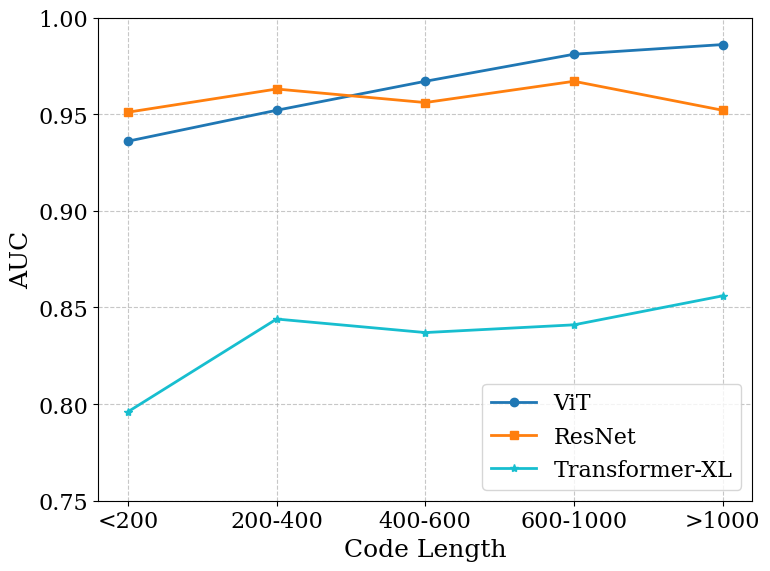}
    \caption{Detection Performance Across Code Lengths}
    \label{fig:len}
\end{figure}

Figure \ref{fig:len} illustrates the effect of code length on the detection performance (AUC) of ViT, ResNet, and Transformer-XL models. ViT outperforms ResNet for medium to long code lengths, while ResNet has a slight edge for shorter snippets. Transformer-XL, while performing lower overall, shows improvement with increasing code length, particularly in the 200-400 token range.

We also conducted experiments on 500 randomly selected code samples to measure time consumption. Our approach involves calling the OpenAI API to compute the log probability distribution, followed by feeding the 2D probability maps into either ResNet or ViT for prediction. As shown in Table \ref{tab:runtime}, the additional time required by ResNet and ViT is negligible compared to the API call. 

The API calling time is significant because it involves remote communication with OpenAI's servers, which introduces network latency and processing delays, especially when dealing with large datasets. This overhead is much higher than the local inference time of ResNet and ViT models, which process the 2D matrices quickly once the log probabilities are obtained. Even with this API overhead, our method still demonstrates significantly lower time consumption compared to perturbation-based methods like DetectGPT and NPR, showcasing its efficiency in real-time applications.

\begin{table}[h]
\centering
\begin{tabular}{lc}
\toprule
\textbf{Method}        & \textbf{Elapsed Time (Sec/Sample)} \\ 
\midrule
DetectGPT            & 29.64                           \\
NPR                  & 31.70                            \\
\midrule
API Calling (pre-processing) & 13.012          \\ 
ResNet                 & 0.000064                        \\ 
ViT                   & 0.000677                        \\ 
\bottomrule
\end{tabular}
\caption{Average elapsed time per sample on a single NVIDIA A100 GPU, comparing our method (API + ResNet/ViT) with DetectGPT and NPR.}
\label{tab:runtime}
\end{table}

\section{Discussion}

\subsection{Limitations}

Our approach relies on the OpenAI API for log probability matrix calculations, which increases cost and makes the method dependent on external service availability, potentially affecting performance and reliability if the API is down or restricted. Additionally, interpretability remains a challenge, as it is difficult to understand why the model makes certain predictions, especially in differentiating human-written from LLM-generated code, limiting its transparency in educational or real-world applications.

\subsection{Future Work}

Future work will focus on replacing the OpenAI API with an open-source model to reduce cost and explore whether smaller language models can effectively generate log probability matrices. We also aim to fully localize the detection process to minimize resource requirements. Another direction involves enhancing interpretability by analyzing which parts of the code are emphasized by attention mechanisms in human vs. AI code. Ablation studies will investigate how the 2D structure impacts ViT's performance by altering code indentation and structure, and we will explore visualization techniques to highlight ViT’s ability to capture code patterns.

\section{Conclusion}

Our method improves upon existing techniques by effectively capturing the hierarchical and syntactical nature of code, leading to more accurate detection of LLM-generated content across multiple programming languages. By training vision models on log probability matrices, we enable the detection of patterns characteristic of human-written and LLM-generated code. This approach combines the strengths of large language models for computing accurate token probabilities and vision models for capturing spatial patterns, offering a powerful and flexible solution. Additionally, our method supports fast detection with minimal computational requirements, making it ideal for deployment on standard classroom computers with stable network connections. This ensures practical and efficient detection of LLM-generated code in educational settings without requiring extensive infrastructure, helping to uphold academic integrity.

\bibliography{aaai25}
\bibliographystyle{aaai}
\end{document}